\documentclass[11pt, a4paper]{article} 

\usepackage[ansinew]{inputenc}
\usepackage{amsmath, amssymb, graphics, amsthm}
\usepackage{epsfig}
\usepackage{color, xcolor}
\usepackage{fancyhdr} 
\usepackage{manfnt}
\usepackage[T1]{fontenc}

\usepackage[normalem]{ulem}

\makeatletter
\@addtoreset{equation}{section}
\makeatother

\newcommand{\be}{\begin{equation}}
\newcommand{\ee}{\end{equation}}
\newcommand{\ba}{\begin{eqnarray}}
\newcommand{\ea}{\end{eqnarray}}

\def\pb#1{\rlap{\lower1.5ex\hbox{$\longleftarrow$}}{#1}}
\def\dpb#1{\rlap{\lower1.5ex\hbox{$\Longleftarrow$}}{#1}}
\def\spb#1{\rlap{\lower1.0ex\hbox{$\leftarrow$}}{#1}}
\def\sdpb#1{\rlap{\lower1.0ex\hbox{$\Leftarrow$}}{#1}}

\DeclareMathOperator{\sign}{sign}

\title{{\sf A note on conformally compactified connection dynamics tailored for anti-de Sitter space}}
\author{
{\sf N. Bodendorfer}\thanks{{\sf 
norbert.bodendorfer@fuw.edu.pl}}\\
\\
{\sf  Faculty of Physics, University of Warsaw, Pasteura 5, 02-093, Warsaw, Poland}\\
}
\date{{\small\sf \today}}

\begin{document} 

\maketitle

{\sf

\begin{abstract}

A framework conceptually based on the conformal techniques employed to study the structure of the gravitational field at infinity is set up in the context of loop quantum gravity to describe asymptotically anti-de Sitter quantum spacetimes. A conformal compactification of the spatial slice is performed, which, in terms of the rescaled metric, has now finite volume, and can thus be conveniently described by spin networks states. The conformal factor used is a physical scalar field, which has the necessary asymptotics for many asymptotically AdS black hole solutions.

\end{abstract}

}

\section{Introduction}

The study of asymptotic conditions on spacetimes has received wide interest within general relativity. Most notably, the concept of asymptotic flatness can serve as a model for isolated systems and one can, e.g., study a system's total energy or the amount of gravitational radiation emitted \cite{WaldGeneralRelativity}. More recently, with the discovery of the AdS/CFT correspondence \cite{MaldacenaTheLargeN}, asymptotically anti de Sitter spacetimes have become a focus of interest. In a certain low energy limit of the correspondence, computations on the gravitational side correspond to solving the Einstein equations with anti-de Sitter asymptotics. 

Within loop quantum gravity (LQG), the issue of non-compact spatial slices or asymptotic conditions on spacetimes has received only little attention so far due to a number of technical reasons. However, also given the recent interest in holography from a loop quantum gravity point of view \cite{BonzomDualityBetweenSpin, Bonzom3DHolography, BIV}, this question is important to address. 
An issue in treating non-compact spatial slices is that cylindrical functions are restricted to depend only on a finite number of holonomies, which prohibits one from writing down quantum states which approximate given canonical data outside of a compact region. This technical problem can however be circumvented by defining such states in the dual of the Hilbert space, which, roughly speaking, allows for spin networks with infinitely many vertices and edges. This approach has been taken by Thiemann in \cite{ThiemannQSD6} to propose a definition for asymptotically flat quantum states and to study the properties of the ADM energy operator thereon. A related strategy to deal with this issue is the infinite tensor product construction \cite{ThiemannGCS4}. 

Another approach taken by Campiglia and Varadarajan \cite{VaradarajanAQuantumKinematics} has been to use the Koslowski-Sahlmann representation \cite{KoslowskiDynamicalQuantumGeometry, SahlmannOnLoopQuantum} of the holonomy-flux algebra, where a background spatial metric (or rather, a densitised triad) is encoded in the quantum state in addition to the cylindrical function. The background spatial metric can then be chosen to satisfy certain asymptotic conditions, while the cylindrical function, encoding excitations above this background, is confined to a compact set. For related work using excitations above a background, see \cite{ArnsdorfLoopQuantumGravity, ArnsdorfLoopQuantumGravityAnd}. 

In this note, we will take a different approach to the subject by employing a conformal compactification of the spatial slice via a rescaling of the canonical variables. In other words, we use a new set of canonical variables, a rescaled metric and its conjugate momentum, such that the proper distance from any bulk point to spatial infinity is finite. 
Quantum states suitable for describing anti-de Sitter space can then be defined directly in the kinematical Hilbert space, that is on finite graphs. Moreover, the usage of such states corresponds to a truncation of theory to a finite number of degrees of freedom, which is of great importance for performing explicit computations.

The classical canonical transformation and the properties of the new variables are the main result of this note and subsequent comments about the quantisation are standard. To avoid confusion, we stress the logic that we only choose new variables for an existing system, general relativity in anti-de Sitter space. In particular, the well-definedness of a given Hamiltonian formulation of this theory in standard metric variables, such as \cite{HenneuaxAsymptoticallyAntiDe}, automatically translates to the new variables. What remains to be checked is the well-definedness of additionally introduced constraints, in our case the Gauss law. In order to ensure a proper transformation under spatial diffeomorphisms, the conformal factor has to be a scalar field modelled in the theory, meaning that we have to add a scalar matter field to our gravity theory.

\section{Connection variables from a conformal compactification}

\subsection{The issue of non-compactness and infinite volume}

Cylindrical functions, the basic elements of the kinematical Hilbert space of LQG, have a finite support on the spatial slice and are thus best suited to describe compact spacetimes, or spacetimes with finite geometric quantities, such as volume. In standard variables, this not the case for anti de-Sitter space. For illustrative purposes, we consider here the Schwarzschild anti-de Sitter black hole given by the metric\footnote{$\Lambda = -3/R^2$ denotes the cosmological constant, $M$ the mass, and $d \Omega^2$ the line element on the $2$-sphere.}
\be
	ds^2 = - \left( 1- \frac{2M}{r} + \frac{r^2}{R^2} \right) dt^2 +  \left( 1- \frac{2M}{r} +  \frac{r^2}{R^2} \right)^{-1} dr^2 + r^2 d \Omega^2 \text{.} \label{eq:SADSBH}
\ee
From \eqref{eq:SADSBH} it follows that the volume of a constant $t$ spatial slice is infinite, as well as the spatial geodesic distance between any bulk point and spatial infinity. Thus, if one were to perform a phase space extension to the usual connection variables derived from the metric and its momentum and follow through with the standard LQG quantisation procedure, one would not be able to approximate the canonical data of anti de-Sitter spacetimes beyond a compact region with spin network states.

\subsection{Compactification of the spatial slice}

A convenient solution to the above problem can be given by applying Penrose' idea of conformal compactification \cite{PenroseAsymptoticPropertiesOf}. As opposed to its standard application, e.g. in anti-de Sitter spacetime as given in \cite{AshtekarAsymptoticallyAntiDe}, we are going to perform a conformal rescaling of the canonical variables, i.e. the spatial metric and its conjugate momentum. After this rescaling, all involved geometric quantities will be finite. We will focus only on the leading order behaviour of the canonical variables, since this is sufficient for the main argument.

We choose a positive conformal factor $\Omega$ which is finite in the bulk and approaches $0$ at spatial infinity ($r \rightarrow \infty$) with a rate of 
\be	
	\Omega \sim \frac{1}{r}  \text{.} \label{eq:ScalarFieldFalloff}  
\ee
We will specify the exact nature of the conformal factor later and apply it first to the problem at hand. We start with the standard canonical variables from the ADM formulation \cite{ArnowittTheDynamicsOf} with the Poisson bracket
\be
	\left\{ q_{ab}(x), P^{cd}(y)\right\} = \delta^{3}(x,y) \delta^{c}_{(a} \delta^d_{b)} \text{.} \label{eq:PBADM}
\ee
Inspired by \cite{AshtekarAsymptoticallyAntiDe}, we perform the rescaling 
\be
	\tilde q_{ab}(x) = \Omega^2(x) q_{ab}(x), ~~~\tilde P^{ab}(x) = \Omega^{-2}(x) P^{ab}(x) \text{.}  \label{eq:RescaledQP}
\ee
It follows that the new twiddled variables still have canonical brackets, 
\be
	\left\{ \tilde q_{ab}(x), \tilde P^{cd}(y)\right\} = \delta^{3}(x,y) \delta^{c}_{(a} \delta^d_{b)} \text{.} \label{eq:PBTilde}
\ee
In particular, the volume of the spatial slices measured with respect to $\tilde q_{ab}$ does not diverge as one is approaching spatial infinity, since\footnote{Given the slicing employed in \eqref{eq:SADSBH}, one still has a divergence at $r = 0$, but this can be cured by employing a different slicing near the black hole, e.g. constant $r$ slices in Gullstrand-Painlev\'e coordinates. We will assume throughout this note that this issue has been dealt with. To keep this in mind, we omit the lower boundary for the r-integral.}
\be
	\int_{S^2} d \Omega \int^{\infty} dr \sqrt{\det \tilde q_{ab}} \sim \int^{\infty} dr  \, r^{-2} = \text{finite} \text{.}
\ee 

\subsection{Connection variables}

From the canonical pair \eqref{eq:PBTilde}, we can construct connection variables along standard lines. An earlier example of a rescaling as in \eqref{eq:RescaledQP} and the subsequent construction of connection variables is provided in \cite{BSTI}, where the motivation was to find a rescaled metric which is invariant with respect to a certain conformal transformation. For simplicity, we will restrict our discussion here to $3+1$ dimensions and construct connection variables analogous to those of Ashtekar and Barbero \cite{AshtekarNewVariablesFor, BarberoRealAshtekarVariables}. First, we perform a phase space extension by introducing the tetrad variables $\tilde K_{ai}, \tilde E^{ai}$, $i,j = 1,2,3$, related to the metric variables as 
\be
	\tilde q \tilde q^{ab} = \tilde E^{ai} \tilde E^{bj} \delta_{ij}, ~~ \tilde P^{ab} = \frac{\sqrt{\det \tilde q}}{2} \left(\tilde K^{ab} -\tilde  q^{ab} \tilde  K \right), ~~ \tilde K_{ab} = \tilde K_{ai} \tilde E^{ci} \tilde q_{bc} / \sqrt{ \det \tilde q}\text{,}
\ee
where spatial tensor indices $a,b$ are raised and lowered with respect to $\tilde q_{ab}$. The three additional gauge degrees of freedom introduced by these variables are taken care of by introducing the Gau{\ss} constraint 
\be
	G_{ij}[\Lambda^{ij}] = 2 \int dr d\Omega \, \Lambda^{ij} \tilde K_{a[i} \tilde E^{a}_{j]} \approx 0 \text{.} \label{eq:GaussLaw}
\ee
Let us now extend the fall-off conditions on the metric and its momentum in anti-de Sitter space as given in \cite{HenneuaxAsymptoticallyAntiDe} to $\tilde E^{ai}$ and $\tilde K_{ai}$. For now, we will only focus on the radial dependence, since this is the essential part for our argument. In particular, we will suppress most angle dependent functions in the following. A straight forward calculation yields for $r \rightarrow \infty$
\be
	\tilde E^{r i} \sim r^0 \, \hat e^{r i}, ~~ \tilde E^{Ai} \sim r^{-2} \hat e^{A i}, ~~ \tilde K_{ri} \sim r^{-3}  \hat e^{ri} + r^{-2} c_B  \hat e^{Bi}, ~~ \tilde K_{Ai} \sim r^{0}  \hat e^{ri} + r^{-1} d_B  \hat e^{Bi} \label{eq:NewVariablesEK}
\ee
where $A,B$ denote angular coordinates orthogonal to $r$ with respect to the asymptotic AdS metric, $\hat e^{ai}$, $a=r,A$, is the internal unit vector in the direction $\tilde E^{ai}$, and $c_B$, $d_B$ are functions of the angles\footnote{We note that while the fall-off conditions on the triad $\tilde E^{a}_i$ are stemming from the leading order behaviour of the AdS metric \eqref{eq:SADSBH}, the fall-off conditions on $\tilde K_{ai}$ stem from perturbations around $\tilde K_{ai} = 0$. In particular, and as opposed to $\tilde E^{ai}$, $\tilde K_{ai}$ can asymptotically vanish faster than indicated in \eqref{eq:NewVariablesEK}. Also, from $\tilde E^{ri}\tilde E^{A}_i = \tilde q \, \tilde q^{rA} \sim r^{-6}$ as $r \rightarrow \infty$, we obtain an additional constraint on $\tilde E^{ai}$.}.
We can allow for $\Lambda^{ij} \sim r^0$ as $r \rightarrow \infty$, so that internal gauge transformations can act non-trivially also at spatial infinity (as opposed to the asymptotically flat case \cite{ThiemannGeneralizedBoundaryConditions, CampigliaNoteOnThe}). 
It also follows from \eqref{eq:NewVariablesEK} that the symplectic structure 
\be
	\Omega (\delta_1, \delta_2) := 2 \int dr d \Omega \, \delta_{[1} \tilde K_{|a|}^i \, \delta_{2]}  \tilde E^{a}_i  \label{eq:SymplStructEK}
\ee
is well defined, i.e. finite, without imposing additional conditions such as parity, again as opposed to the asymptotically flat case \cite{ReggeRoleOfSurface, BeigThePoincareGroup}.

It is very instructive to compute the fall-off conditions in a new coordinate system which is adapted to our idea of describing the whole spatial slice with a finite range of coordinates after the conformal compactification. To this end, let us choose a coordinate $x$ instead of $r$ which for $r\rightarrow \infty$ behaves as $r = \frac{x}{1-x} \Leftrightarrow x = \frac{r}{1+r}$. Clearly, $x \rightarrow 1$ from below as $r\rightarrow \infty$. In these new coordinates, the fall-off conditions read
\be
	\tilde E^{x i} \sim x^0 \, \hat e^{x i}, ~~ \tilde E^{Ai} \sim x^{0} \hat e^{A i}, ~~ \tilde K_{xi} \sim (1-x)  \hat e^{xi} + x^{0} c_B  \hat e^{Bi}, ~~ \tilde K_{Ai} \sim x^{0}  \hat e^{xi} + (1-x) d_B  \hat e^{Bi} \text{,}\label{eq:NewVariablesEKinX}
\ee
which makes the finiteness e.g. \eqref{eq:GaussLaw} or \eqref{eq:SymplStructEK} manifest. Also, it is manifest now that twiddled geometric quantities are asymptotically finite since they are obtained from integrations of finite densities over finite regions. The conformal factor behaves as $\Omega \sim \frac{1-x}{x}$ as $x \rightarrow 1$. In particular, we see that $\partial_x \Omega  |_{x=1} \sim 1$, in accordance with the usual definition \cite{AshtekarAsymptoticallyAntiDe}.

In a last step, we perform a canonical transformation to connection variables $\tilde A_{ai}, \tilde E^{ai}$ by adding the spin connection $\tilde \Gamma_a^i(\tilde E^{ai})$, to $\tilde K_{a}^i$ \footnote{For simplicity, we neglect the possible introduction of the Barbero-Immirzi parameter in this note.}, see e.g. \cite{ThiemannModernCanonicalQuantum}. 
We extend the fall-off conditions on $\tilde K_{ai}$ to $\tilde A_{ai}$ by demanding that $\tilde A_{a}^i - \tilde \Gamma_{a}^i(\tilde E)$ falls off as $K_{a}^i$ in \eqref{eq:NewVariablesEK}, instead of specifying some fall-off conditions directly on $A_{a}^i$.
We note that this condition is trivially preserved under the evolution generated by the Gau{\ss} law, since both $\tilde A_{a}^i$ and $\tilde \Gamma_{a}^i$ transform as connections. It is also preserved by the Hamiltonian and spatial diffeomorphism constraints due to the general properties of a canonical transformation.

To check canonicity of the transformation to connection variables, the relation\footnote{Assuming $\sign \det \tilde e = 1$ for simplicity.}
\be 
	\tilde E^{a}_i \delta \tilde \Gamma_a^i = - \frac{1}{2} \partial_a \left( \epsilon^{abc} \delta (\tilde e_b^j) \tilde e_{cj} \right) \text{,}
\ee
see e.g. \cite{ThiemannModernCanonicalQuantum}, is key. Using it, the symplectic structure \eqref{eq:SymplStructEK} can simply be rewritten as\be
	\Omega (\delta_1, \delta_2) = 2 \int_\Sigma dr d \Omega \, \delta_{[1} \tilde K_{|a|}^i \, \delta_{2]}  \tilde E^{a}_i = 2 \int_\Sigma dr d \Omega \, \delta_{[1} \tilde A_{|a|}^i \, \delta_{2]}  \tilde E^{a}_i +  \int_{\partial \Sigma} d \Omega \, \epsilon^{AB} \delta_{[1} \tilde e_{|A|}^j \delta_{2]} \tilde e_{Bj} \text{.}
\ee
It follows that $A_a^i$ and $E^a_i$ are canonically conjugate and that a boundary contribution to the symplectic structure emerges.
In particular, since $\tilde e_{A}^i \sim 1$ as $r \rightarrow \infty$, the boundary contribution to the symplectic structure is finite and thus also the bulk contribution. 
Following the results of \cite{AshtekarIsolatedHorizonsThe, EngleBlackHoleEntropy, BII}, the boundary symplectic structure can further be expressed as the one of Chern-Simons theory and quantised accordingly \cite{EngleBlackHoleEntropy, AshtekarQuantumGeometryOf}. 

\subsection{Choice of conformal factor}

In order to apply standard LQG techniques to connection variables derived from the twiddled variables \eqref{eq:RescaledQP}, we need to ensure that they transform properly under spatial diffeomorphisms, which is not possible with a randomly picked $\Omega$. 
Rather, one needs to use a conformal factor which transforms as a scalar field under the spatial diffeomorphism constraint of the theory. An example is provided by the following construction:

A wide range of black hole solutions with anti-de Sitter asymptotics is available for minimally and non-minimally coupled scalar fields, see e.g. \cite{MartinezTopologicalBlackHoles} and references therein. While some details of these solutions differ, their unifying property is that the scalar field is non-vanishing and falls off at spatial infinity as \eqref{eq:ScalarFieldFalloff} at leading order, whereas the metric asymptotically approaches \eqref{eq:SADSBH}. We can thus directly use such a scalar field, denoted by $\psi$, as a conformal factor $\Omega$.
We rewrite the symplectic potential as
\be
	\int_\Sigma d^3x \, \left( P^{ab} \delta q_{ab} + \pi_\psi \delta \psi \right) = \int_\Sigma d^3x \, \left(\tilde  P^{ab} \delta \tilde q_{ab} + \left( \pi_\psi - \frac{2 P^{ab} q_{ab}}{\psi}\right) \delta \psi \right) \text{,} \label{eq:TwiddledWithScalarField}
\ee
and thus define $\tilde \pi_\psi := \pi_\psi - \frac{2 P^{ab} q_{ab}}{\psi}$. The canonical pair $\{ \psi(x), \tilde \pi_\psi(y)\} = \delta^{(3)}(x,y)$ can now be treated as a scalar field which commutes with the twiddled variables.

\section{Comments}

Quantisation can now proceed along standard lines \cite{ThiemannModernCanonicalQuantum, RovelliQuantumGravity}, since our system is just general relativity coupled to a scalar field written using connection variables satisfying all properties necessary for loop quantum gravity techniques to apply. 
The resulting spin networks however correspond to quanta of geometry as measured by the twiddled variables \eqref{eq:RescaledQP} as in \cite{BSTI}. As an example, the corresponding area operator would have its usual quantised spectrum \cite{AshtekarQuantumTheoryOf1}, however measuring $\widetilde {\text{Area}} := \text{Area} \cdot \Omega^2$. A similar statement is true for the volume operator. 
Implementing the bulk part of the constraints follows standard lines. The spatial diffeomorphism constraint is solved by the methods of \cite{AshtekarQuantizationOfDiffeomorphism}. For the Hamiltonian constraint, we need to rewrite the constraints using our connection variables and apply the techniques of \cite{ThiemannQSD1, ThiemannQSD5}. No problems are expected here for the bulk part of the constraint, since the techniques of \cite{ThiemannQSD1} have already been applied to more complicated systems \cite{MaNonperturbativeLoopQuantization}. A strategy for the quantisation of the boundary part of the Hamiltonian constraint is left for future research.

As a representation for the scalar field, it would be interesting to consider the proposal of \cite{LewandowskiLoopQuantumGravityCoupled}, where $\psi(x)$ can be sharply peaked everywhere on the spatial slice. However, such a representation is hard to incorporate into the framework of \cite{ThiemannQSD1, ThiemannQSD5, ThiemannKinematicalHilbertSpaces}, roughly because also the spatial derivative of $\psi$ acts as a local operator and one cannot absorb all the powers of $\epsilon$ in a regularisation of the Hamiltonian constraint. We thus choose the representation defined in \cite{ThiemannQSD5, ThiemannKinematicalHilbertSpaces, AshtekarPolymerAndFock}, which does not encounter this difficulty.

In this note, we have not commented on how exactly the asymptotic conditions for anti-de Sitter space should be implemented on the quantum states. In principle, one could choose a certain graph, representing a certain truncation on the spatial slice, and construct coherent states thereon following \cite{ThiemannGCS1}. A key point of interest thereby is to understand how to properly extract the asymptotic symmetries of anti-de Sitter spacetime. 

The main difference between our work and \cite{VaradarajanAQuantumKinematics, KoslowskiDynamicalQuantumGeometry, SahlmannOnLoopQuantum, ArnsdorfLoopQuantumGravity, ArnsdorfLoopQuantumGravityAnd} is that we do not use any background state above which we consider a compact set of excitations. Up to the different variables, we work in the standard Ashtekar-Isham-Lewandowski representation.  We also do not resort to a definition in the dual of the Hilbert space as in \cite{ThiemannQSD6}. In all of those works, the quanta encoded by spin networks correspond to geometry, while in our case, due to the rescaling of the metric, they correspond to a combination of geometry and the scalar field.

\section{Conclusion}

We have pointed out that after a conformal compactification of the spatial slice, one can construct loop quantum gravity in such a way that slices of anti-de Sitter spacetime can be approximated by regular spin networks. 
An important motivation behind this construction is that finite truncations to the relevant degrees of freedom, in this case the restriction to work on a given graph, are important for practical calculations. 

The main motivation behind this research is to apply loop quantum gravity techniques to supergravity in anti-de Sitter space in order to use the results to compute properties of gauge theories via the AdS/CFT correspondence \cite{BTTVIII, BIV}. Quantum gravity techniques are expected to be necessary here to extend these computations away from the large $N$ regime which is described by the classical gravity approximation usually employed. Success along these lines could thus have significant impact. 

It would also be interesting to extend the results of this note to the asymptotically flat context or to find different ways to incorporate conformal factors, possibly via a modified radial gauge fixing \cite{DuchObservablesForGeneral, BLSII} such as $\Omega^2 q_{ra} = \delta_{ra}$.

\section*{Acknowledgements}
NB was supported by a Feodor Lynen Research Fellowship of the Alexander von Humboldt-Foundation and during final improvements of this manuscript by the Polish National Science Centre grant No. 2012/05/E/ST2/03308 as well as a Feodor Lynen Return Fellowship.


\end{document}